%% file: luszczak.tex
\documentclass[12pt]{article}
\usepackage{graphicx}

\newcommand\pubdate{\today}

\def\napoli{Faculty of Mathematics and Natural Sciences, University of Rzesz\'ow,
ul.  Pigonia 1, 35-310 Rzesz\'ow \\
Institute of Nuclear Physics PAN, PL-31-342 Krak\'ow, Poland}
\def\support{\footnote{luszczak@ur.edu.pl}}

\def\Title#1{\begin{center} {\Large #1 } \end{center}}
\def\Author#1{\begin{center}{ \sc #1} \end{center}}
\def\Address#1{\begin{center}{ \it #1} \end{center}}

\newcommand\pubblock{\rightline{\begin{tabular}{l} \\
         \pubdate  \end{tabular}}}
\newenvironment{Abstract}{\begin{quotation}  }{\end{quotation}}
\newenvironment{Presented}{\begin{quotation} \begin{center} 
             PRESENTED AT\end{center}\bigskip 
      \begin{center}\begin{large}}{\end{large}\end{center} \end{quotation}}
\def\Acknowledgements{\bigskip  \bigskip \begin{center} \begin{large}
             \bf ACKNOWLEDGEMENTS \end{large}\end{center}}

\input econfmacros.tex

\begin{document}
\begin{titlepage}
\pubblock

\vfill
\Title{Single-diffractive dijet production at high energies within the $k_t$ -factorization approach}
\vfill
\Author{Marta Luszczak\support and Antoni Szczurek}
\Address{\napoli}
\vfill
\begin{Abstract}
We discuss single-diffractive production of dijets. The cross section is
calculated for the first time in 
the $k_t$-factorization approach, neglecting transverse momentum of the pomeron.
We use Kimber-Martin-Ryskin unintegrated parton (gluon, quark,
antiquark) distributions (UPDF) both in the proton as well as in the
pomeron or subleading reggeon.  The UPDFs are calculated based on 
conventional MMHT2014nlo PDFs in the proton and  H1 collaboration 
diffractive PDFs used previously in the analysis of diffractive 
structure function and dijets at HERA.  
We try to describe the existing data from Tevatron
and make detailed predictions for possible LHC measurements. 

\end{Abstract}
\vfill
\begin{Presented}
EDS Blois 2017, Prague, \\ Czech Republic, June 26-30, 2017
\end{Presented}
\vfill
\end{titlepage}
\def\thefootnote{\fnsymbol{footnote}}
\setcounter{footnote}{0}

\section{Introduction}

We discuss single-diffractive production of
dijets. This process was discussed in the past for photo- and
electro-production as well as for
proton-proton or proton-antiproton collisions.
The hard single diffractive processes are treated usually in the resolved pomeron
picture with a pomeron being a virtual but composed (of partons)
object. This picture was used with a success for the description of 
hard diffractive processes studied extensively at HERA.
This picture was tried to be used also for hadronic collisions. 
A few processes were studied experimentally at the Tevatron including the dijet production.
We propose the single-diffractive dijet 
production for the first time within the $k_t$-factorization approach,
see \cite{Luszczak:2017pna}.
Similar approach was used recently for the  single-diffractive
production of $c \bar c$ pairs \cite{Luszczak:2016csq}.
In particular, we wish to compare results 
obtained within collinear-factorization and $k_t$-factorization
approaches. A comparison with the Tevatron data and predictions for the LHC
are presented. 

\section{A sketch of the approach}

\begin{figure}[htb]
\centering
\includegraphics[height=1.95in]{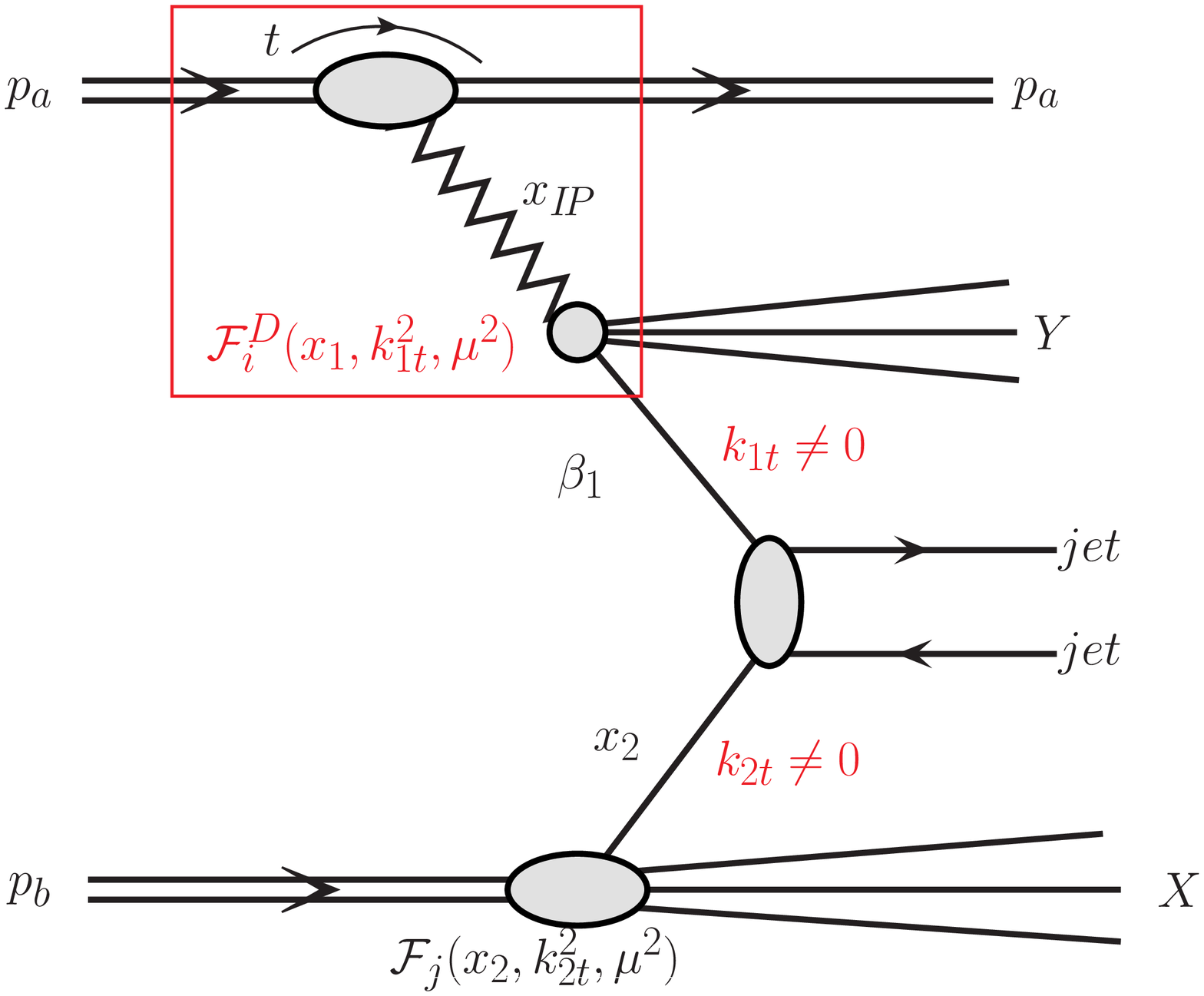}
\includegraphics[height=1.95in]{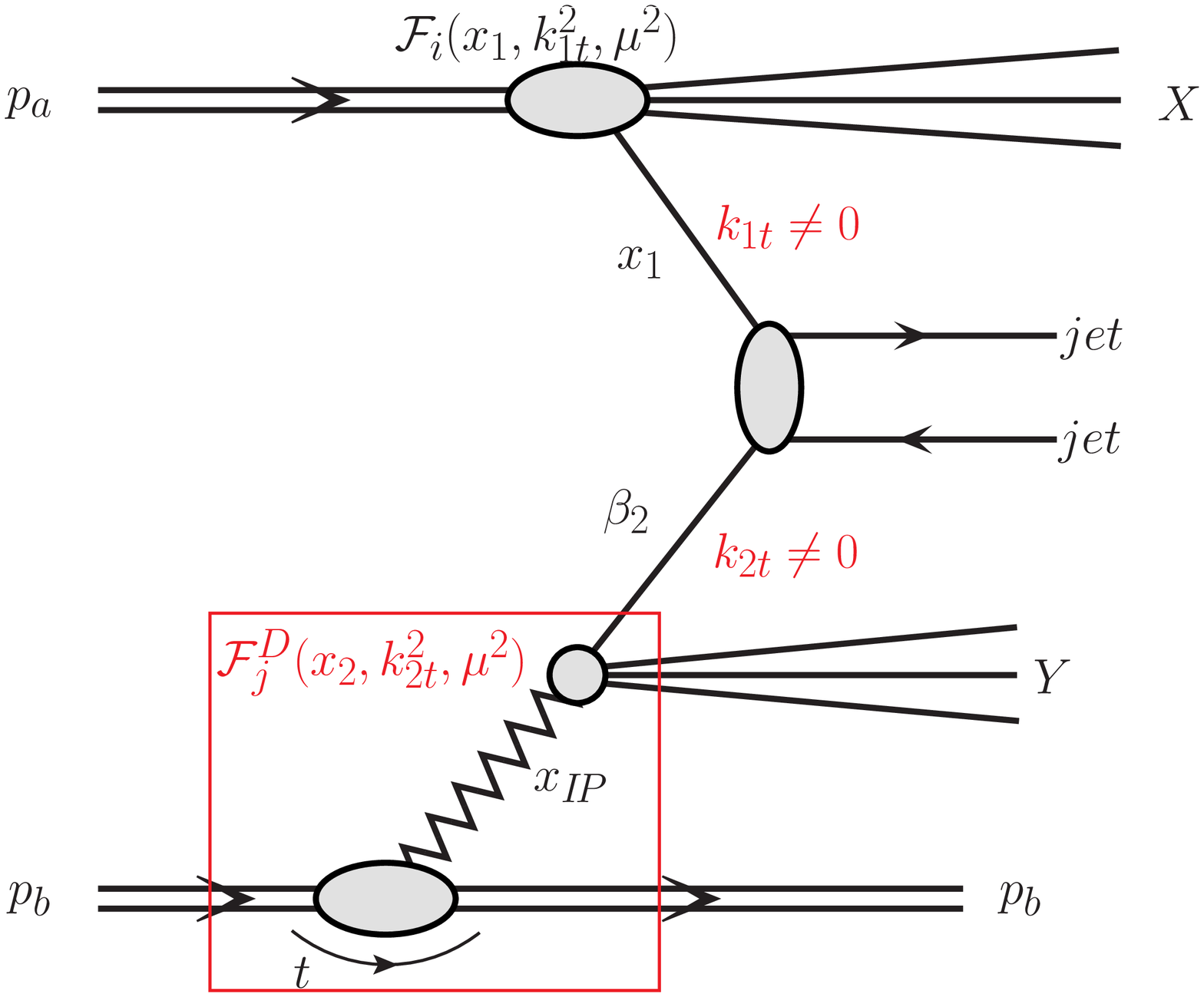}
\caption{A diagrammatic representation of the considered mechanisms of single-diffractive dijet production.}
\label{fig:mechanism}
\end{figure}

According to the approach sketched above in Fig.~\ref{fig:mechanism}, the cross section for inclusive single-diffractive production of dijet, for both considered diagrams (left and right panel of Fig.~\ref{fig:mechanism}), can be written as:
\begin{eqnarray}
d \sigma^{SD(1)}({p_{a} p_{b} \to p_{a} \; \mathrm{dijet} \; X Y}) &=& \sum_{i,j,k,l}
\int d x_1 \frac{d^2 k_{1t}}{\pi} d x_2 \frac{d^2 k_{2t}}{\pi} \; d {\hat \sigma}({i^{*}j^{*} \to kl }) \nonumber \\
&& \times \; {\cal F}_{i}^{D}(x_1,k_{1t}^2,\mu^2) \cdot {\cal F}_{j}(x_2,k_{2t}^2,\mu^2) ,
\label{SDA_formula}
\end{eqnarray}
\begin{eqnarray}
d \sigma^{SD(2)}({p_{a} p_{b} \to \mathrm{dijet} \; p_{b} \; X Y}) &=& \sum_{i,j,k,l}
\int d x_1 \frac{d^2 k_{1t}}{\pi} d x_2 \frac{d^2 k_{2t}}{\pi} \; d {\hat \sigma}({i^{*}j^{*} \to kl }) \nonumber \\
&& \times \; {\cal F}_{i}(x_1,k_{1t}^2,\mu^2) \cdot {\cal F}_{j}^{D}(x_2,k_{2t}^2,\mu^2),
\label{SDB_formula}
\end{eqnarray}
where ${\cal F}_{i}(x,k_{t}^2,\mu^2)$ are the "conventional" unintegrated ($k_{t}$-dependent) parton distributions (UPDFs) in the proton and ${\cal F}_{i}^{D}(x,k_{t}^2,\mu^2)$ are their diffractive counterparts -- which we will call here diffractive UPDFs (dUPDFs). Details of our new calculations can be found in \cite{Luszczak:2017pna}.

\section{Results}
\begin{figure}[!htbp]
\begin{minipage}{0.47\textwidth}
 \centerline{\includegraphics[width=1.0\textwidth]{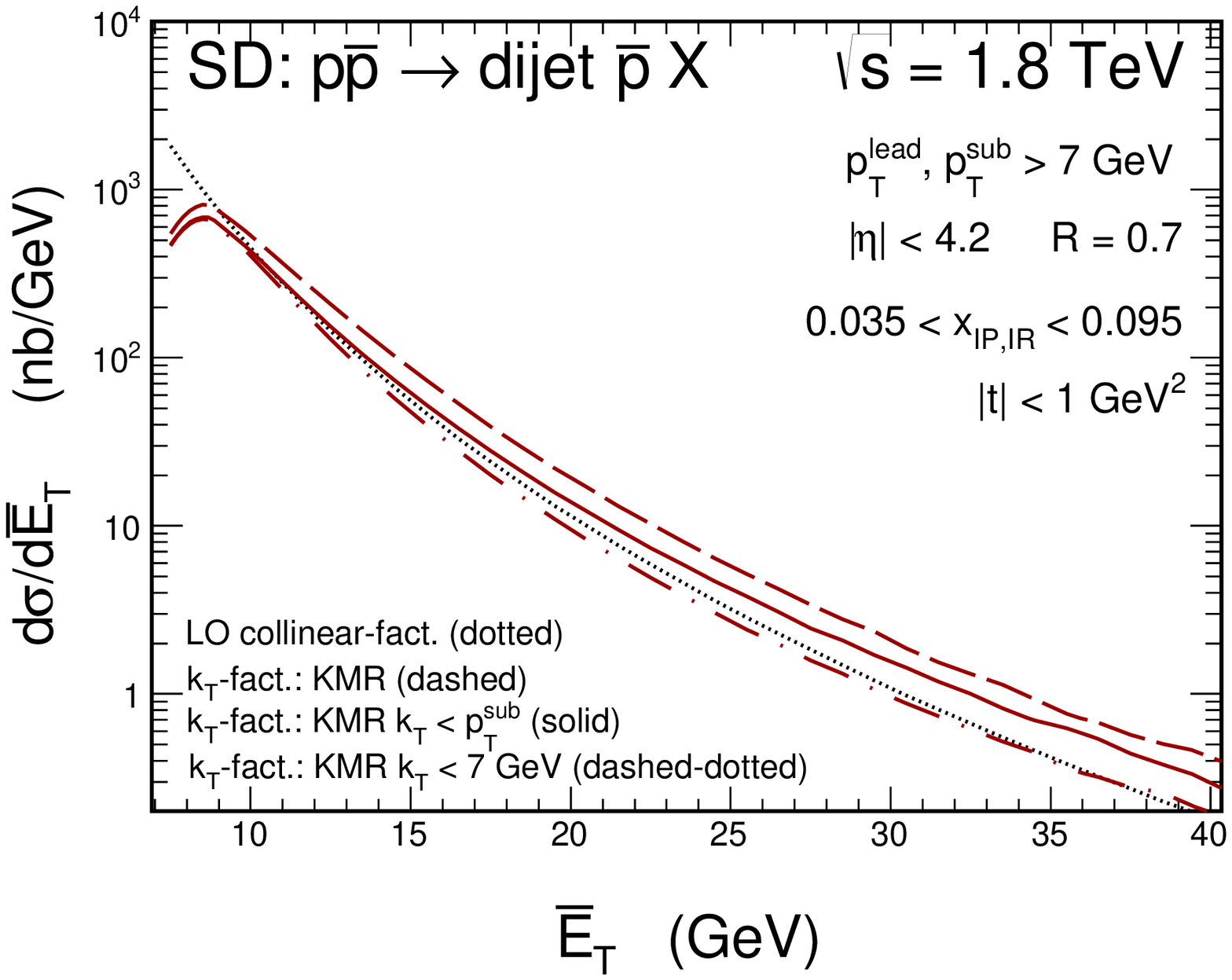}}
\end{minipage}
\hspace{0.5cm}
\begin{minipage}{0.47\textwidth}
 \centerline{\includegraphics[width=1.0\textwidth]{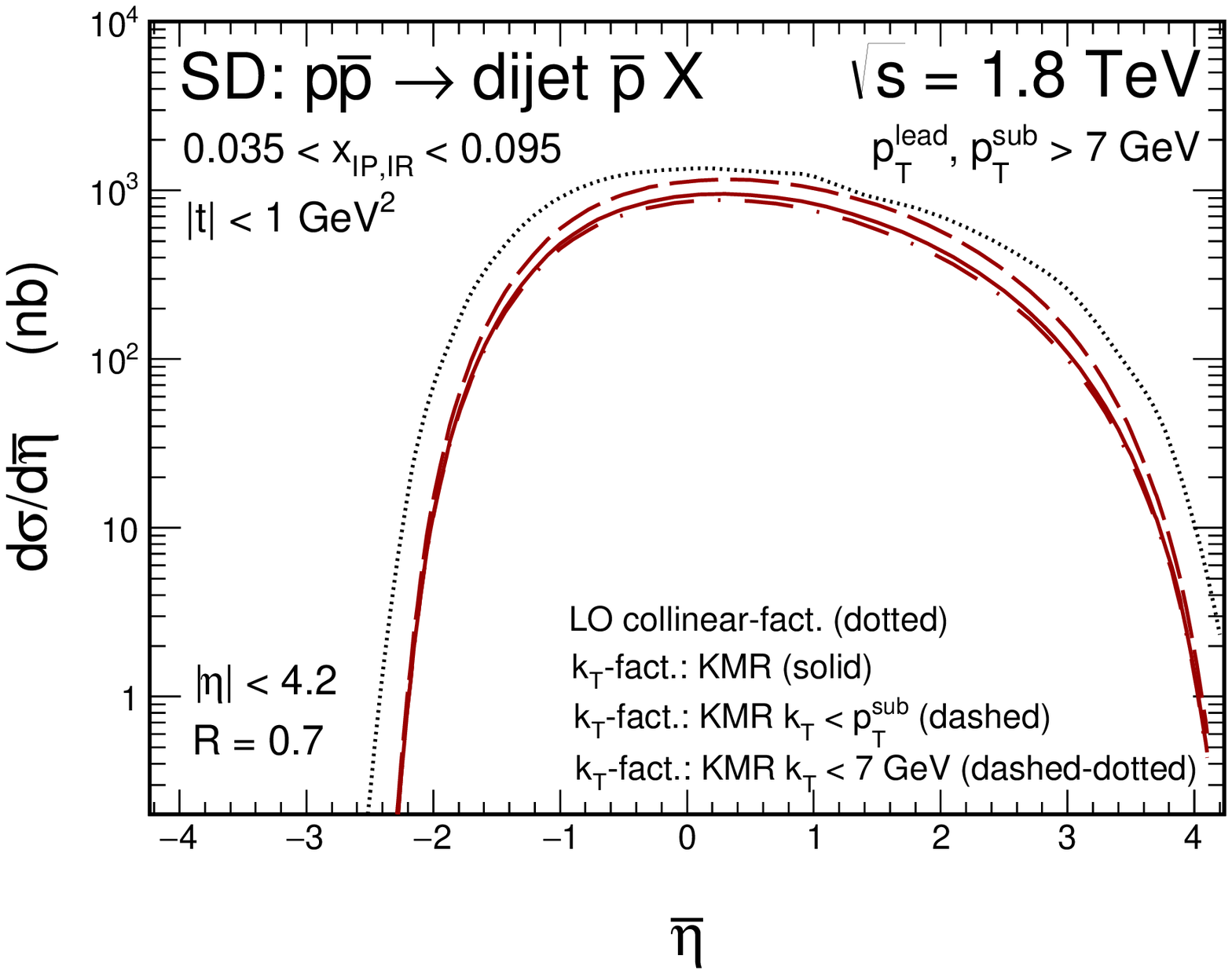}}
\end{minipage}
   \caption{
\small Distribution in average ${\overline E}_T$ (left panel) and
in average ${\overline \eta}$ (right panel). Here $S_G$ = 0.1 was assumed.
}
 \label{fig:dsig-1800_SD}
\end{figure}

\begin{figure}[!htbp]
\begin{minipage}{0.47\textwidth}
 \centerline{\includegraphics[width=1.0\textwidth]{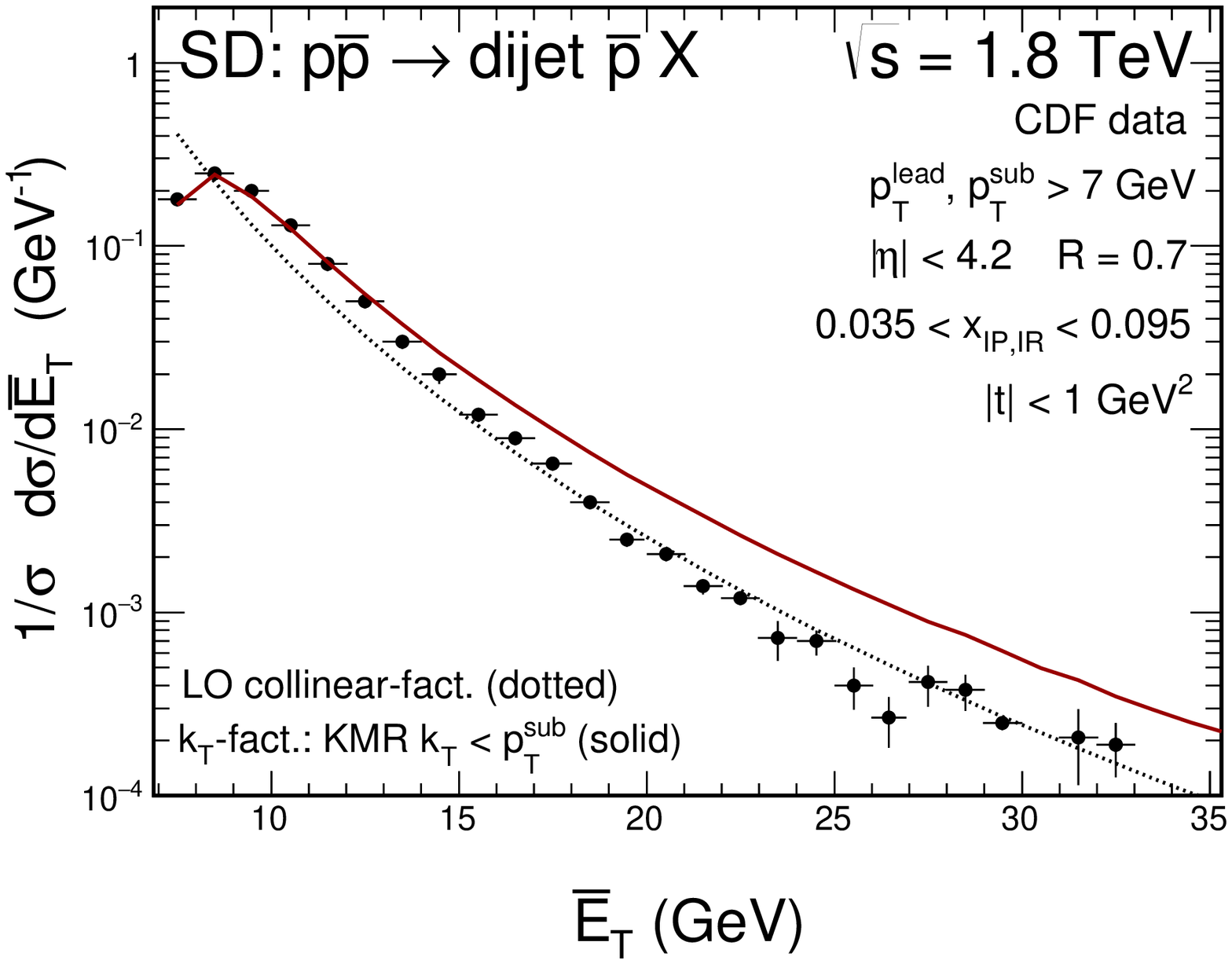}}
\end{minipage}
\hspace{0.5cm}
\begin{minipage}{0.47\textwidth}
 \centerline{\includegraphics[width=1.0\textwidth]{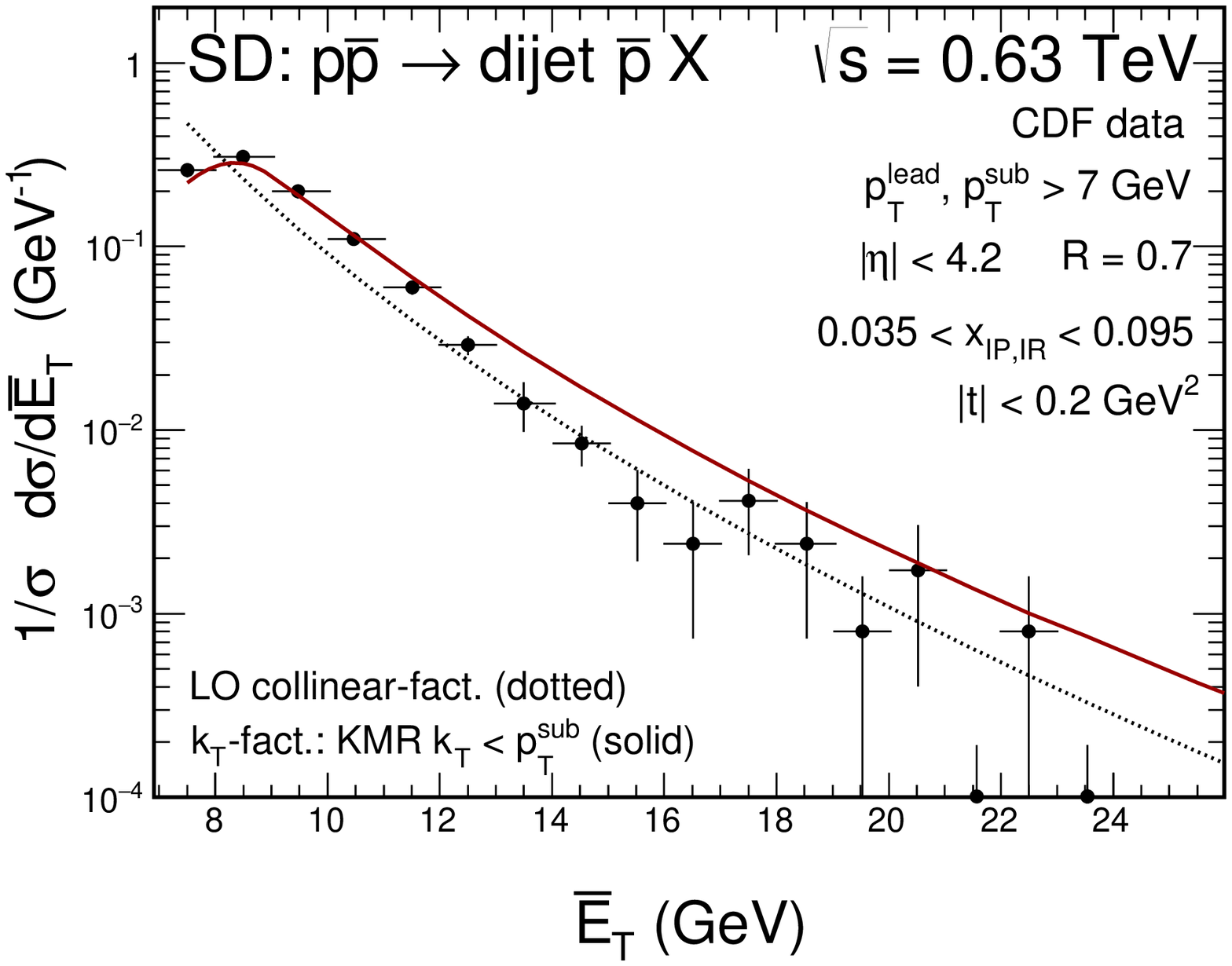}}
\end{minipage}
   \caption{
\small The average transverse energy distribution for $\sqrt{s}$ = 1.8 TeV
(left panel) and for $\sqrt{s}$ = 630 GeV (right panel).
}
 \label{fig:cdf_pt}
\end{figure}

We start by showing our results for ${\overline E}_T = \frac{E_{1T} + E_{2T}}{2}$
and ${\overline \eta} = \frac{\eta_1 + \eta_2}{2}$ distributions, see Fig.~\ref{fig:dsig-1800_SD}.
In this calculation the pomeron/reggeon longitudinal momentum 
fraction was limited as in experimental case \cite{Affolder:2000vb,Affolder:2001zn} 
to 0.035 $< x_{I\!P,I\!R} <$ 0.095.
We show both naive result obtained with the KMR UGDF (dashed line)
as well as similar results with limitations on parton transverse
momenta $k_T < p_T^{sub}$ (solid line) and $k_T < 7 $ TeV (dash-dotted line). Above $p_T^{sub}$ is transverse momentum of the subleading jet.
The first limitation was proposed for standard nondiffractive jets
\cite{Nefedov:2013ywa}. 
A large difference can be seen close to the lower transverse momentum
cut.
In Fig.~\ref{fig:cdf_pt} we show distribution in ${\overline E}_T$ for two collision
energies. While the $k_t$-factorization approach gives a better
description of the data close to the lower experimental cut on
jet transverse momenta, the collinear-factorization approach seems to
be better for larger transverse momenta.

\begin{figure}[!htbp]
\begin{minipage}{0.47\textwidth}
 \centerline{\includegraphics[width=1.0\textwidth]{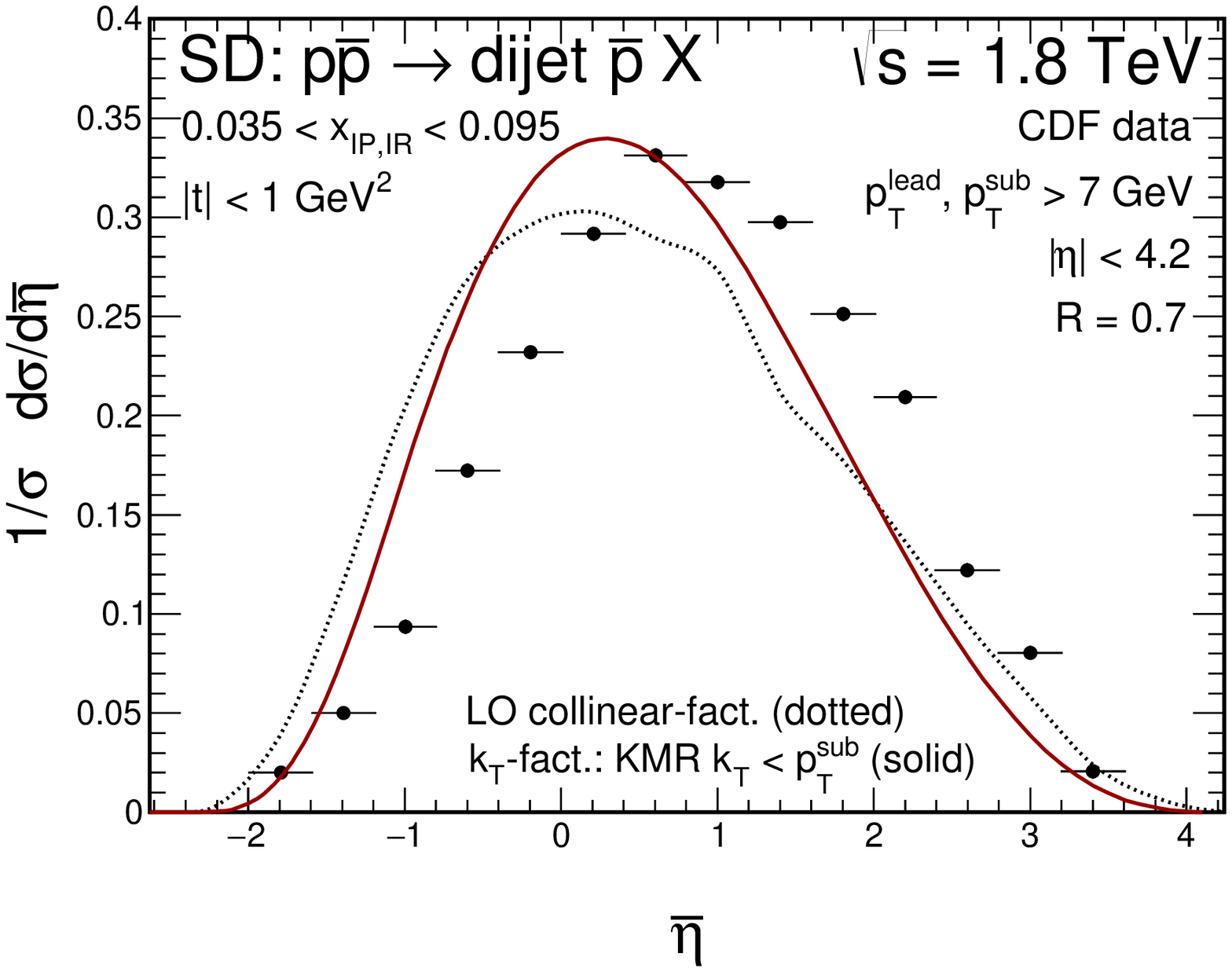}}
\end{minipage}
\hspace{0.5cm}
\begin{minipage}{0.47\textwidth}
 \centerline{\includegraphics[width=1.0\textwidth]{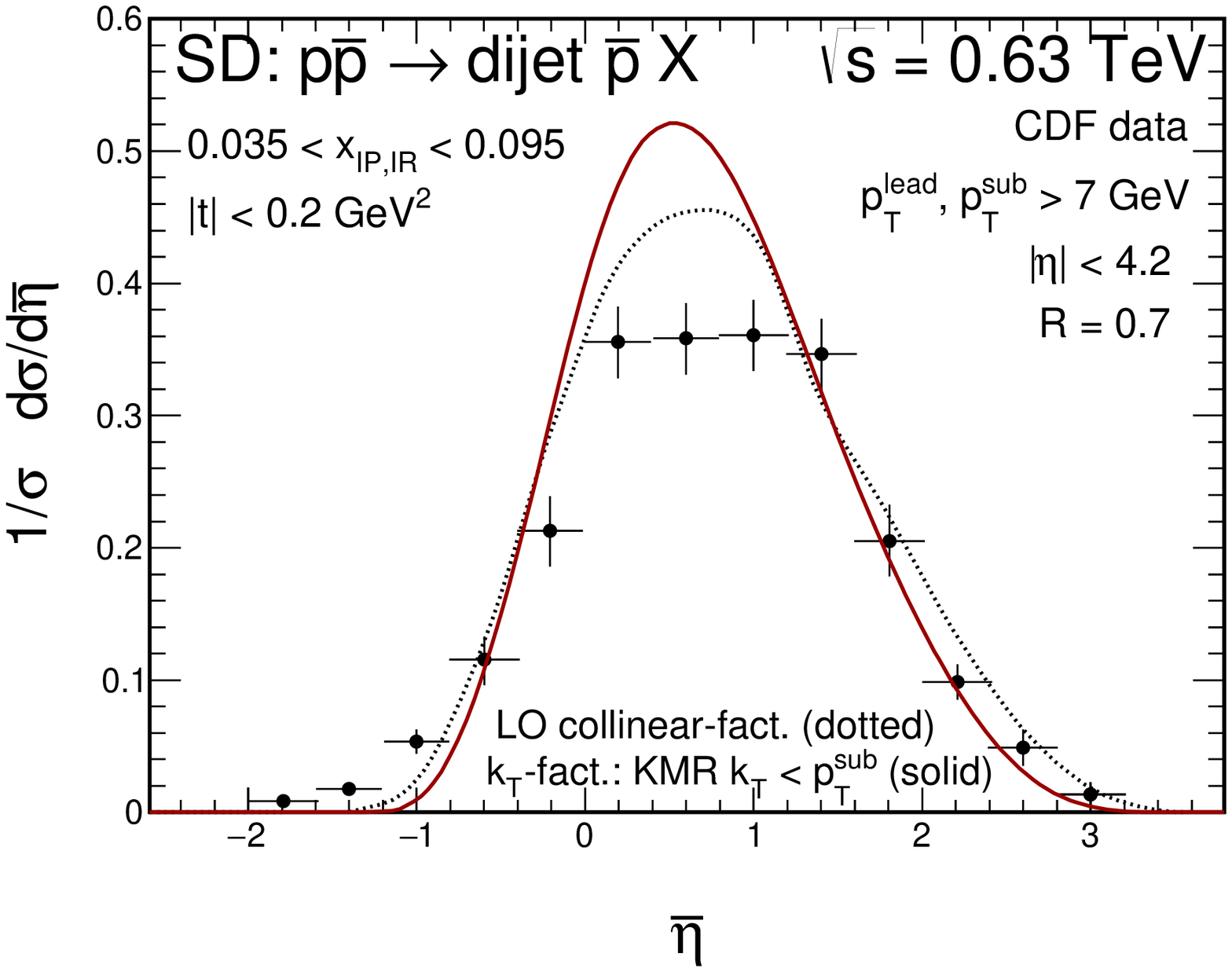}}
\end{minipage}
   \caption{
\small The average rapidity dsistribution for $\sqrt{s}$ = 1.8 TeV
(left panel) and for $\sqrt{s}$ = 630 GeV (right panel).
}
 \label{fig:cdf_y}
\end{figure}

In Fig.~\ref{fig:cdf_y} we show distributions in average jet rapidity again for 
the two collision energies. Here the $k_t$-factorization result 
better describes the experimental data than the result obtained in 
the collinear approach. The outgoing antiproton is at
$\eta \approx$ -6.05 for $\sqrt{s}$ = 1.8 TeV and $\eta \approx$ -5.53 for $\sqrt{s}$ = 630 GeV,
respectively.

\begin{figure}[!htbp]
\begin{minipage}{0.47\textwidth}
 \centerline{\includegraphics[width=1.0\textwidth]{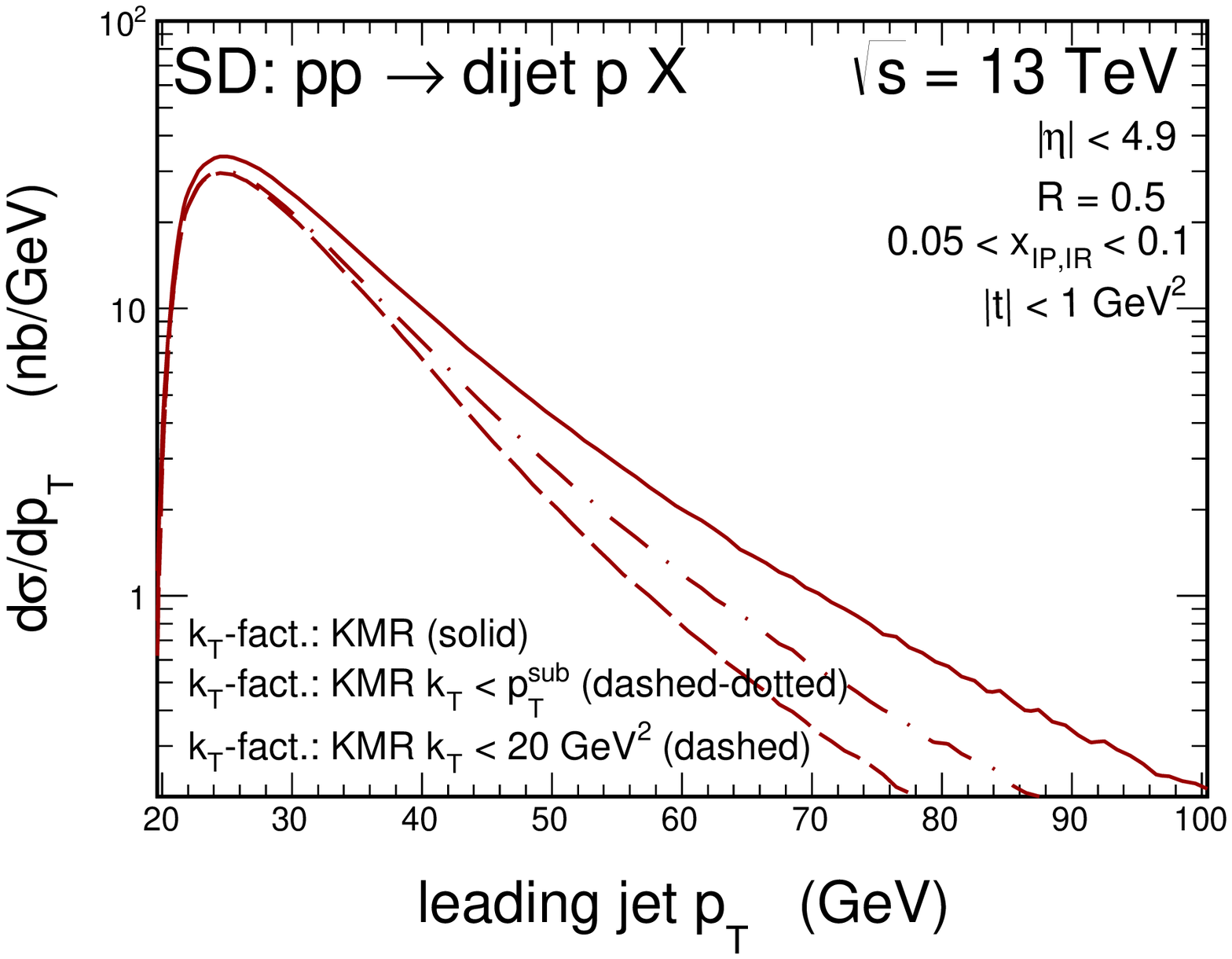}}
\end{minipage}
\hspace{0.5cm}
\begin{minipage}{0.47\textwidth}
 \centerline{\includegraphics[width=1.0\textwidth]{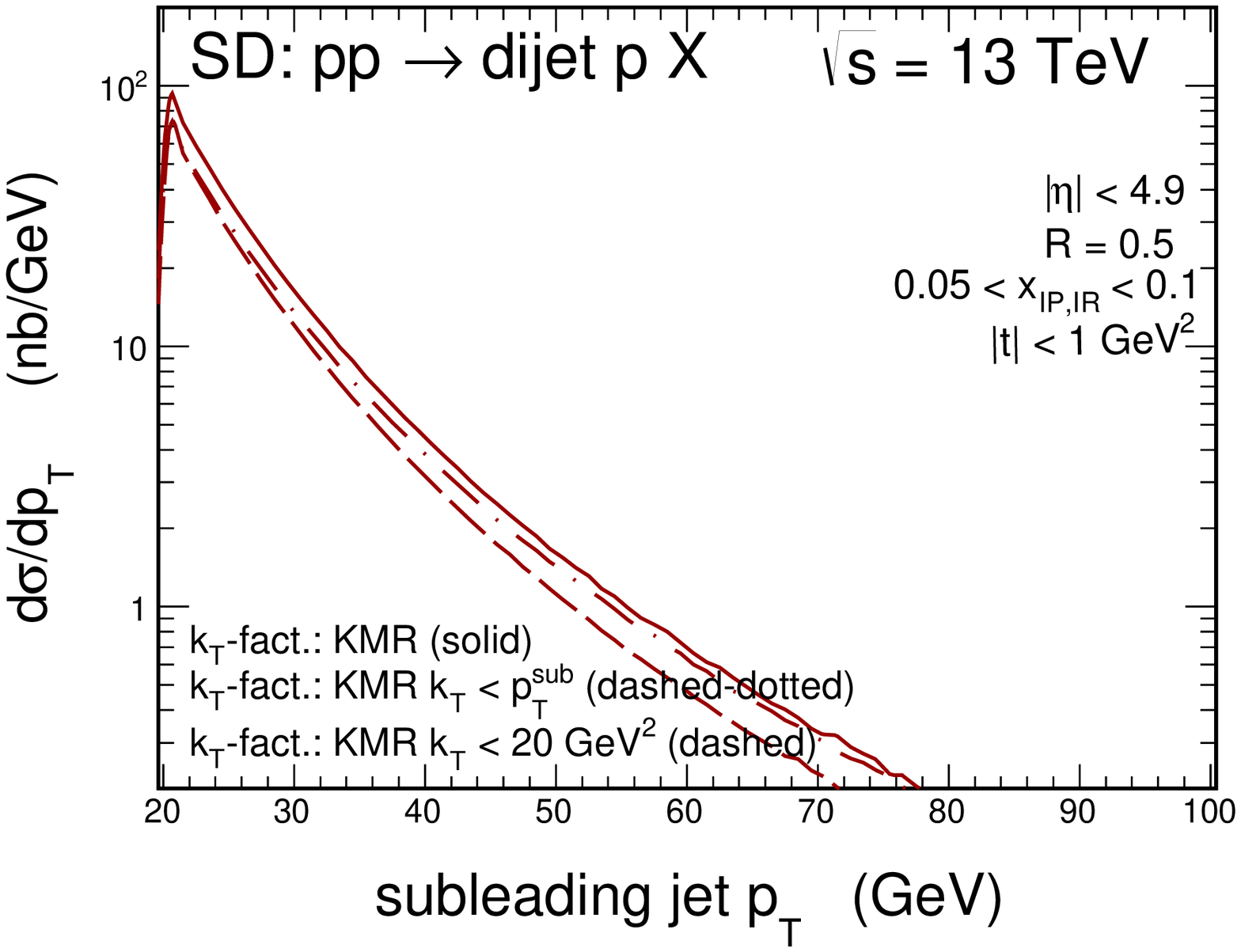}}
\end{minipage}
   \caption{
\small Distribution in the jet transverse momentum for leading
(left panel) and subleading (right panel) for $\sqrt{s}$ = 13 TeV
and for the ATLAS cuts. Here $S_G$ = 0.05.
}
 \label{fig:7}
\end{figure}

In Fig.~\ref{fig:7} we show distribution in jet transverse momentum,
for leading (left panel) and subjeading (right panel) jets.
As for the Tevatron we discuss the role of extra cuts on parton 
transverse momenta. The cuts have bigger efect on leading jets.

\section{Conclusion}
We have presented results for
the single-diffractive production of dijets within $k_t$-factorization
approach. Results of our calculations were compared with the Tevatron data 
where forward antiprotons and rapidity gaps were measured. We have 
calculated distributions in ${\overline{E}_T}$ and ${\overline \eta}$. 
A resonable agreement has been achieved.
We have compared results obtained within collinear and 
$k_t$-factorization approaches.
The $k_t$-factorization leads to a better description in 
$E_T$ close to the lower transverse momentum cut.
\Acknowledgements
This study was partially supported by the Polish National Science Centre grant DEC-2013/09/D/ST2/03724 and by the Center for Innovation and Transfer of
Natural Sciences and Engineering Knowledge in Rzesz\'ow.

\end{document}

%% file: econfmacros.tex
\def\beq{\begin{equation}}
\def\eeq#1{\label{#1}\end{equation}}
\def\eeqn{\end{equation}}

\def\beqa{\begin{eqnarray}}
\def\eeqa#1{\label{#1}\end{eqnarray}}
\def\eeqan{\end{eqnarray}}

\let\bar=\overbar

\def\Dslash{\not{\hbox{\kern-4pt $D$}}}
\def\dslash{\not{\hbox{\kern-2pt $\del$}}}

\def\msb{{\bar{\ssstyle M \kern -1pt S}}}

